\documentclass[aps,preprint,nofootinbib]{revtex4}
\usepackage{graphicx}
\usepackage{amsmath,amssymb,amscd}
\usepackage[caption=false]{subfig}

\newcommand*{\ra}{\rightarrow}

\newcommand*{\lra}{\longrightarrow}

\newcommand*{\del}{\delta}

\newcommand*{\gm}{\gamma}
\newcommand*{\half}{\frac{1}{2}}

\DeclareMathOperator{\Tr}{Tr}

\begin{document}
\title{Constraints on New Scalars from the LHC 125 GeV Higgs Signal}

\author{We-Fu Chang}
\affiliation{Department of Physics, National Tsing Hua University, and Physics Division, National Center for Theoretical Sciences,
 No. 101, Section 2, Kuang-Fu Road, Hsinchu, Taiwan 30013, R.O.C.}
\email{wfchang@phys.nthu.edu.tw}
\author{John  N. Ng}
\affiliation{TRIUMF, 4004 Wesbrook Mall, Vancouver, BC, V6T 2A3, Canada}
\email{misery@triumf.ca}
\author{Jackson M. S. Wu}
\affiliation{Physics Division, National Center for Theoretical Sciences, No. 101, Section 2, Kuang-Fu Road, Hsinchu, Taiwan 30013, R.O.C.}
\email{jknw350@yahoo.com}

\begin{abstract}
We study the implications the recent results from the LHC Higgs
searches have on scalar new physics. We study the impact on both
the Higgs production and decay from scalars with and without
color, and in cases where decoupling do and do not happen. We
investigate possible constraints on scalar parameters from the
production rate in the diphoton channel, and also the two vector
boson channels. Measurements from both channels can help
disentangle new physics due to color from that due to charge, and
thus reveal the nature of the new scalar states.
\end{abstract}

\maketitle

\section{Introduction}
A central piece of the Large Hadron Collider (LHC) programme is to find the Standard Model (SM) Higgs boson, or else discover scalar particle(s) of similar properties.
The LHC has great sensitivity for Higgs produced via gluon fusion, which then decays into electroweak gauge bosons. So far, no significant excess of events has been seen in the
mass range of 129~GeV to 600~GeV at 95\% confidence level~\cite{LHCcon}. But at around 125~GeV, both the ATLAS~\cite{Atlas1} and CMS~\cite{CMS1} detectors observed an excess at the
$3\sigma$ level in the diphoton channel. The ATLAS collaboration also reported an excess in the $ZZ^*$ channel in this mass range~\cite{Atlas2}, although the CMS search in this
channel yielded a less significant result~\cite{CMS2}. Taken at face value, the recent LHC data seems to suggest a central value of the diphoton production rate 1.5 to 2 times the
SM value, but one consistent with the SM for the $ZZ^*$ production. There were also Higgs searches at the Tevatron, which is sensitive for Higgs mass below 200~GeV. A broad excess
interpretable as a SM Higgs decaying into a pair of bottom quarks was observed~\cite{Teva}. The statistical significance is not sufficient to claim discovery. Nevertheless,
it is tantalizing. The current run at the LHC will certainly clarify the situation.

The unexpected enhancement in the diphoton production rate has
motivated many recent studies on possible new physics (NP)
explanations (see e.g.~\cite{many} and references within). The
diphoton channel is a very clean channel that is sensitive to
corrections in both the production cross-section and the decay
width. New scalar states have been widely considered as possible
sources. A lot of studies have investigated in specific models
their impact on either the Higgs production via gluon-gluon fusion
(GF), which is the dominant production process, or the Higgs decay
in the diphoton channel. Here, we study the generic effect the new
scalars have on both the Higgs production and decay. Besides the
GF production and the diphoton decay, we also include in our study
the productions from vector bosons fusion (VBF) and associate
production with vector boson (VH), $V = W,Z$, and decay into two
vector bosons. We emphasize that by measuring the production rates
in different decay channels, or in the same decay channel but from
different production processes, information about the color and
the charge of the new scalar states can be
obtained~\footnote{Currently, the associate production cannot be
separated out from GF and other production mechanism. However, we
anticipate this to be feasible in the future LHC runs with high
energy and high luminosity.}.

A feature we pay close attention to is whether the effects of the
new scalars states decouple as they become heavy. In general,
colored scalars are naturally of the decoupling type as their
masses do not share the same origin as the SM Higgs. On the other
hand, scalars that mixes with the SM Higgs tend not to decouple.
Properties of the new scalar state may be revealed from this
perspective. This has not been emphasized before, and it could be
a useful tool for indirect searches of NP.

This paper is organized as follows. In Section~\ref{sec:general}
we discuss in general how NP can modify the LHC Higgs signal. In
section~\ref{sec:scalarNP} we study classes of scalar NP according
to their $SU(3)$ color and $SU(2)$ weak representations using
explicit examples, and we investigate the constraints from
production rates in the $\gm\gm$ and $VV^*$ channel.
Section~\ref{sec:concl} contains our conclusions.

\section{\label{sec:general} General Considerations}
At the LHC, the Higgs boson is produced predominantly through GF. To a lesser degree production also proceeds through VBF; VH processes can also be non-negligible if the Higgs is light.
For the decay, we focus mainly on the diphoton and the $VV^*$ channels where the LHC has sensitivity for. In the SM, both the GF production of
the Higgs and its diphoton decay start at the one-loop level, while VBF, VH, and $H \ra VV^*$ are all processes that start at the tree level. Most NP scenarios do not alter the tree level processes.
An exception is when there is large admixture of new spin-0 states with the SM Higgs, or new spin-1 states with the SM gauge bosons. Since no new vector bosons were found, it is reasonable to assume
that the as yet undiscovered spin-1 states have small mixing with the SM gauge bosons. The question of the Higgs mixing with new scalar or pseudoscalar states is however open. In general, such mixing will reduce the strength of the Higgs couplings to $W$ and $Z$ with respect to their SM values. The Higgs couplings to fermions are likely to be similarly affected which, in particular, would have consequence
for $H \ra b\bar{b}$, the dominant decay mode for a light 125~GeV Higgs. In later sections we study the effects of scalar mixings with explicit examples.

We focus here on how NP affect the loop-induced processes: $gg \ra
H$ and $H\ra\gm\gm$. To have our analysis as model independent as
possible, we base our study on the spin, masses and couplings of
the NP. We discuss the generalities of NP effects from spin-0 and
spin-1/2 new states below. We leave the investigation of higher
spin NP for future works.

In order to affect the GF process, the new degrees of freedom must
couple to the Higgs and have non-trivial color $SU(3)$
representations. Similarly, to affect the diphoton decay they must
be electrically charged. Thus new neutrino states, sterile or
otherwise, have no bearing in our consideration. We note that
since the NP would interfere with SM processes at the amplitude
level, it is not surprising that if confirmed, the LHC signal can
severely restrict the possible couplings of these new degrees of
freedoms with the SM Higgs.

Consider first the Higgs diphoton decay. Including spin-0 and spin-1/2 NP contributions, the width is given by
\begin{align}\label{eq:2gamma}
\Gamma_{\gm\gm}\equiv\Gamma(H\ra\gm\gm) &= \frac{G_\mu \alpha^2 M_H^3}{128\sqrt{2}\pi^3}\Bigg|F_1(\tau_W) + \frac{4}{3}F_{1/2}(\tau_t) \notag \\
&\qquad +
\sum_{\phi}d(r_\phi)Q_\phi^2\frac{\lambda_\phi}{g_w}\frac{M_W^2}{M_{\phi}^2}F_0(\tau_\phi)
+ \sum_{f}d(r_f)Q_f^2\frac{2
y_f}{g_w}\frac{M_W}{M_f}F_{1/2}(\tau_f)\Bigg|^2 \,,
\end{align}
where $G_\mu$ is the Fermi constant, $\alpha$ the fine-structure constant, and $g_w$ the weak gauge coupling. The first two terms are the SM one-loop contributions from the $W$ and the top quark. It is well known that the $W$-loop dominates in the SM and the top contribution subtracts from it. Following the conventions of Ref.~\cite{Dj08}, we define $\tau_i \equiv M_H^2/(4M_i^2)$, and the one-loop functions are given by
\begin{align}\label{eq:Functions}
F_0(\tau) &= -[\tau - f(\tau)]\tau^{-2} \,, \notag \\
F_{1/2}(\tau) &= 2\left[\tau + (\tau - 1)f(\tau)\right]\tau^{-2} \,, \notag \\
F_1(\tau) &= -\left[2\tau^2 + 3\tau + 3(2\tau - 1)f(\tau)\right]\tau^{-2} \,,
\end{align}
with
\begin{equation}
f(\tau) =
\begin{cases}
\arcsin^2\sqrt{\tau} & \tau \leq 1 \\
-\frac{1}{4}\left[\log\frac{1 + \sqrt{1 - \tau^{-1}}}{1 - \sqrt{1 - \tau^{-1}}} - i\pi\right]^2  & \tau > 1
\end{cases} \,.
\end{equation}

In NP contributions, we see explicitly the dependence on $Q$, the
electric charge of the new particle. The color dependence enters
through $d(r)$, the dimension of the color representation $r$ of
the new particle. Unlike the Yukawa coupling, $y_f$, of the Higgs
to new fermions, the $H\phi\phi$ coupling is dimensionful. For
convenience we scale it using the $W$ boson mass so that it is
$\lambda_\phi M_W$. In general the phases of $y_f$ and
$\lambda_\phi$ relative to $g_w$ are not determined. However, they
can be fixed in specific models.

For the GF production, the parton level cross section can be written as
\begin{equation}\label{eq:sigg}
\sigma_{gg}\equiv\hat{\sigma}(gg \ra H) = \sigma_0\,M_H^2\,\del(\hat{s} - M_H^2) \,,
\end{equation}
where
\begin{equation}\label{eq:sig0}
\sigma_0 = \frac{G_\mu\alpha_s^2}{128\sqrt{2}\pi}\left|\half
F_{1/2}(\tau_t) +
\sum_{\phi}\frac{\lambda_\phi}{g_w}\frac{M_W^2}{M_\phi^2}C(r_\phi)F_0(\tau_\phi)
+ \sum_{f}\frac{2
y_f}{g_w}\frac{M_W}{M_f}C(r_f)F_{1/2}(\tau_f)\right|^2 \,.
\end{equation}
The first term is the SM contribution dominated by the top loop.
The color dependence in the NP contributions enters here through
$C(r)$, the index of the representation $r$~\footnote{For a
particle in the $SU(3)$ representation $r$ with generator $T^a_r$,
$\Tr T^a_r T^b_r = C(r)\del^{ab}$. Normalizing such that
$C(\mathbf{3}) = 1/2$, we have $C(\mathbf{6})= 5/2$ and
$C(\mathbf{8})=3$.}.

Comparing Eq.~\eqref{eq:2gamma} and~\eqref{eq:sig0}, we see
similar NP contributions to both the GF production and the
diphoton decay. The differences are in the color factors and
electric charges. Partial compensation of NP contributions are
thus expected when considering the event rates from the reaction
chain $gg \ra H \ra \gm\gm$. This has been noted previously in the
study of models with 4th generation fermions. 

From Eq.~\eqref{eq:2gamma} and~\eqref{eq:sig0}, we see also that NP contributions scales as $M_\phi^{-2}$ for scalars and $M_F^{-1}$ for fermions for large masses. This follows from the fact
that in the large mass limit
\begin{equation}
F_0(\tau) \xrightarrow[\tau \ra 0]{} \frac{1}{3} \,, \qquad F_{1/2}(\tau) \xrightarrow[\tau \ra 0]{} \frac{4}{3} \,.
\end{equation}
It may appear at first sight that NP would always decouple as new
particles become heavy. However, this is not necessary the case in
general: decoupling would not happen if the effective couplings
$\lambda_\phi$ and $y_f$ also scale as the mass. This is
illustrated for example by 4th generation models with a SM Higgs
doublet, where Yukawa couplings $y_f \sim  M_F/M_W$. The mass
dependence resides in $F_{1/2}(\tau)$, which approaches a
constant. The non-decoupling effect leads to the well known demise
of the model as noted in~\cite{CMS4}. The LHC Higgs search in
channels considered above are particularly sensitive for this
class of models.

In the following we specialize to scalar NP. We pay particular
attention to cases where non-decoupling occurs. All our analyses
below are performed for the case of LHC at $\sqrt{s} = 8$~TeV.

\section{\label{sec:scalarNP} Scalar New Physics}
\subsection{Scalars with color: general features}
If a scalar field $\phi$ carries color, its mass cannot be the
result of spontaneous symmetry breaking (SSB): a colored vacuum
would result otherwise. Prominent classes of examples are squarks
in supersymmetric models, scalar leptoquarks that couple to quarks
and leptons, and fermiophobic colored scalars. The squark masses
arise mainly from soft breaking mass and can be taken to infinity
independent of electroweak symmetry breaking. Similarly, masses of
the leptoquark and colored scalar are free parameters in the
effective theory at or near the electroweak scale. The only
relevant coupling is the $H\phi^\dagger \phi$ coupling, which can
arise from the gauge invariant interaction term
\begin{equation}
\lambda_\phi \, \phi^\dagger \phi H^\dagger H \overset{SSB}{\lra}\lambda_\phi\frac{v}{\sqrt{2}}\phi^\dagger\phi H \,,
\end{equation}
where $v = 246$~GeV sets the Fermi scale, and SSB contributes an
amount $\lambda_\phi v^2/2$ to the mass of the scalar, $M^2_\phi$.
The effective coupling $\lambda_\phi$ is independent of $M_\phi$,
and such scalars are of the decoupling type. The sign of
$\lambda_\phi$ is not fixed \textit{\`{a} priori} and is model
dependent. Colored scalars can contribute to either or both GF
production and the diphoton decay. VBF and VH productions are not
affected.

\begin{figure}[htbp]
\centering
\includegraphics[width=5in]{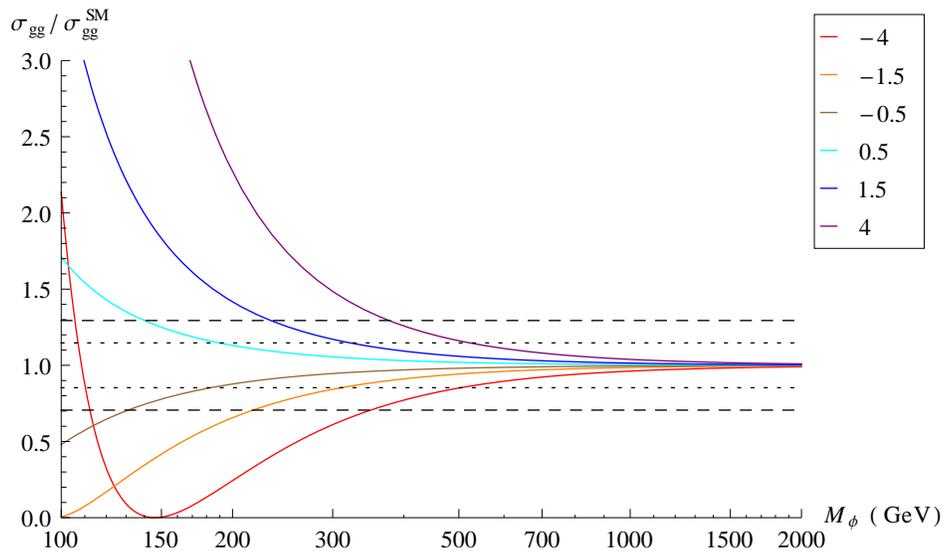}
\caption{The SM normalized Higgs production cross-section via GF
for various values of $\lambda_\phi C(r_\phi)$ as denoted by the
colored lines. The dotted (dashed) lines give bounds from the
theoretical $1\sigma$ ($2\sigma$) uncertainty in the determination
of $\sigma_{gg}^{SM}$.} \label{fig:rxsGF}
\end{figure}

\begin{figure}[htbp]
\centering
\includegraphics[width=5in]{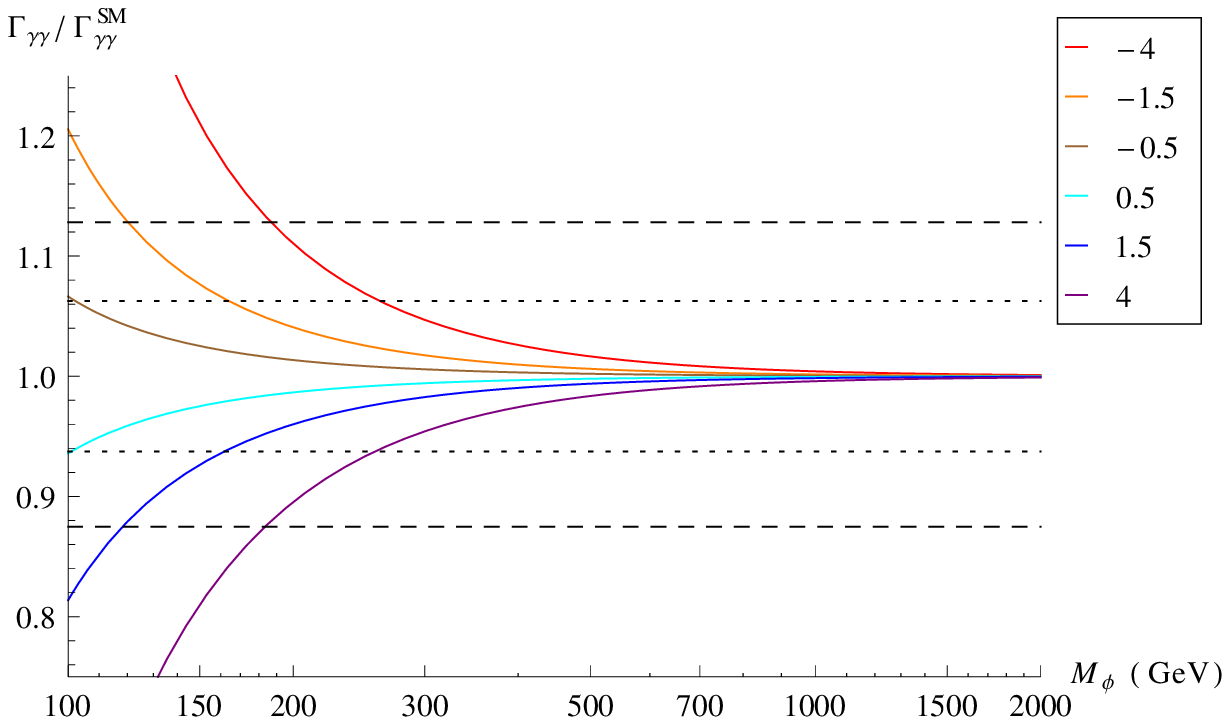}
\caption{The SM normalized Higgs diphoton decay width for various
values of $\lambda_\phi d(r_\phi)Q_\phi^2$ as denoted by the
colored lines. The dotted (dashed) lines give bounds from the
theoretical $1\sigma$ ($2\sigma$) uncertainty in the determination
of $\Gamma_{\gm\gm}^{SM}$.} \label{fig:rdwaa}
\end{figure}
We illustrate in Figs.~\ref{fig:rxsGF} and~\ref{fig:rdwaa} how the
GF production cross-section and the diphoton decay width of the
Higgs are changed with respect to the SM by the colored scalar NP.
The decoupling nature of the colored scalar is evident. It sets in
at $M_\phi \sim 1$~Tev for the range of coupling we consider.
Comparing the two figures, we see also how the partial
compensation of NP effects can happen from production to decay.
Noting that the current data have very low statistics, we use the
expected SM values as a guide for constraints on the NP parameter
space. Specifically, we take the theoretical $1\sigma$ uncertainty
in the SM normalization used as a benchmark. Here, the theoretical
$1\sigma$ uncertainty in $\sigma_{gg}^{SM}$ is $\pm 14.7\%$, and
in $\Gamma_{\gm\gm}^{SM}$ $[-6.3\%,6.4\%]$~\cite{LHCHXSWG}.

As can be seen from Eq.~\eqref{eq:2gamma} and~\eqref{eq:sig0}, for
$\lambda_\phi$ sufficiently negative, there can be a cancellation
between the SM top and the scalar NP contributions. Moreover, for
$M_\phi$ sufficiently light, the scalar NP contribution can become
large and completely dominate over that from the SM top. Such
behavior is seen in Fig.~\ref{fig:rxsGF} in the case of
$\lambda\,C(r_\phi) = -4$, where $\sigma_{gg} = 0$ at $M_\phi =
146.1$~GeV (for $\lambda\,C(r_\phi) = -1.5$, the zero occurs at
$M_S = 97.0$~GeV).

Recent LHC data suggests that $R_{\gm\gm} \sim 1.5$ to 2~\cite{Atlas1,CMS1}, where $R_{\gm\gm}$ is the ratio of the diphoton production rates defined by
\begin{equation}\label{eq:R2photons}
R_{\gm\gm} = \frac{\sigma(pp \ra H + X)\mathrm{Br}(H \ra \gm\gm)}{\sigma_{SM}(pp \ra H + X)\mathrm{Br}_{SM}(H \ra \gm\gm)} \,.
\end{equation}
Given that GF is the dominant Higgs production mechanism at the
LHC (about 88\% of the total cross-section~\cite{LHCHXSWG}), and
that branching ratio is not expected to be wildly different from
the SM for most mass ranges (cf. Fig.~\ref{fig:rdwaa}), such
enhancement should be reflected in the ratio
$\sigma_{gg}/\sigma_{gg}^{SM}$. Fig.~\ref{fig:rxsGF} would then
suggest that having $\lambda_\phi$ negative ($C(r_\phi)$ is
positive) is unlikely to give rise to the purported enhancement,
unless its magnitude is large. However, the allowed range for
$M_\phi$ is then much smaller than for the corresponding postil
valued coupling. This is again illustrated by the
$\lambda\,C(r_\phi) = -4$ case.

Although at the LHC, Higgs production cross-sections and decay
widths (or branching ratios) are not separately measured,
information pertaining to the color and the charge of the new
scalar states can be extracted by simultaneously measuring both
$R_{\gm\gm}$ and $R_{VV}$, a quantity similarly defined but for
the $H \ra VV^*$ channel. This is so because the two have
parametrically different dependence on $d(r_\phi)Q_\phi^2$ in the
branching ratios.

Such procedure can be made even sharper if VH production
cross-section can be measured on its own, separate from that of
the GF, which is the only production mechanism affected by the
color of the NP. This is anticipated to be possible at 14~TeV with
high luminosity and high statistics. By measuring also
\begin{equation}\label{eq:GH2photon}
R^{VH}_{\gm\gm} = \frac{\sigma(pp \ra VH)\mathrm{Br}(H\ra\gm\gm)}{\sigma_{SM}(pp \ra VH)\mathrm{Br}_{SM}(H\ra\gm\gm)} \,,
\end{equation}
complementary information on the color representation can be
extracted, which can then be used to extract information about the
charge $Q_\phi$.

Below we study in detail the role of the electric charge in
colored scalars.

\subsubsection{Electrically neutral colored scalars: $Q_\phi = 0$}
By being electrically neutral, these colored scalars contribute
only to the GF process. The GF production cross-section can be
altered significantly for large color representations. All other
Higgs production channel and the diphoton decay are unchanged from
the SM.

\begin{figure}
\centering
\includegraphics[width=5in]{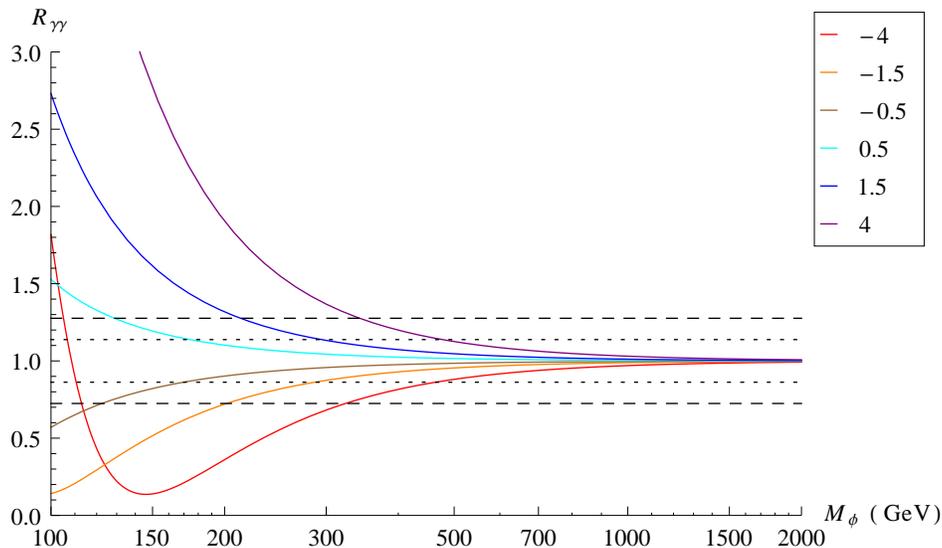}
\caption{The SM normalized production rate $R_{\gm\gm}$ for the
case of charge neutral scalar NP. The colored lines denote various
values of $\lambda_\phi C(r_\phi)$. The dotted (dashed) lines give
bounds from the total theoretical $1\sigma$ ($2\sigma$)
uncertainties in the SM diphoton production rate.}
\label{fig:RaaQ0}
\end{figure}
We show in Fig.~\ref{fig:RaaQ0} the ratio $R_{\gm\gm}$ as a
function of $M_\phi$, the mass of the colored scalar particle, for
various values of $\lambda_\phi C(r_\phi)$. The theoretical
$1\sigma$ uncertainty in the SM diphoton production rate is $\pm
13.8\%$. In making the plot and deriving the uncertainties, we
have used values of SM Higgs production cross-sections and
branching ratios compiled in Ref.~\cite{LHCHXSWG}.

Note for $Q_\phi = 0$, $R_{VV} = R_{\gm\gm}$. This is an important
prediction for this class of scalar NP. Preliminary data suggests
that the two are not equal~\cite{LHCcon}. If this finding
persists, it would rule out the electrically neutral colored
scalars as the source of deviation from the SM.

\subsubsection{Electrically charged colored scalars: $Q_\phi \neq 0$}
With electric charge, the colored scalars modify both the GF
production cross-section and the diphoton decay width.
\begin{figure}
\centering
\subfloat[$|\lambda_\phi C(r_\phi)| = 4$]{
\includegraphics[width=3in]{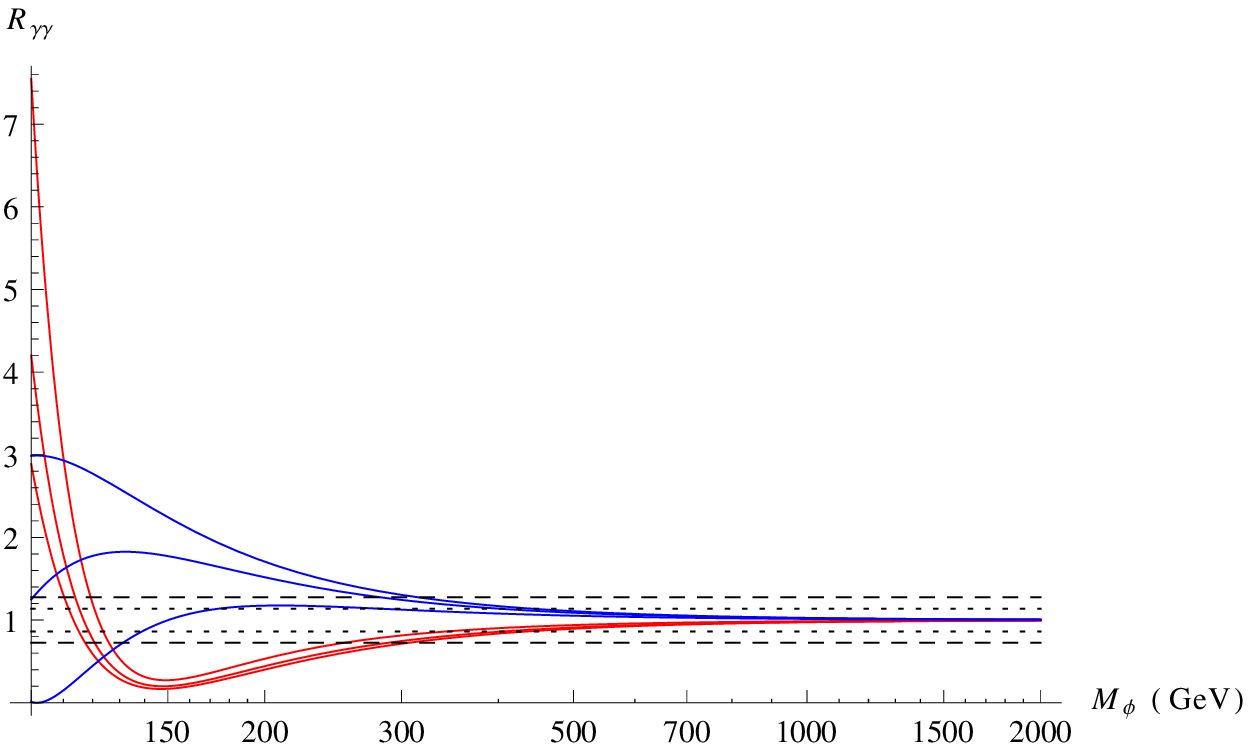}
\label{fig:subfig:QlC4}}
\hspace{0.2in}
\subfloat[$|\lambda_\phi C(r_\phi)| = 2$]{
\includegraphics[width=3in]{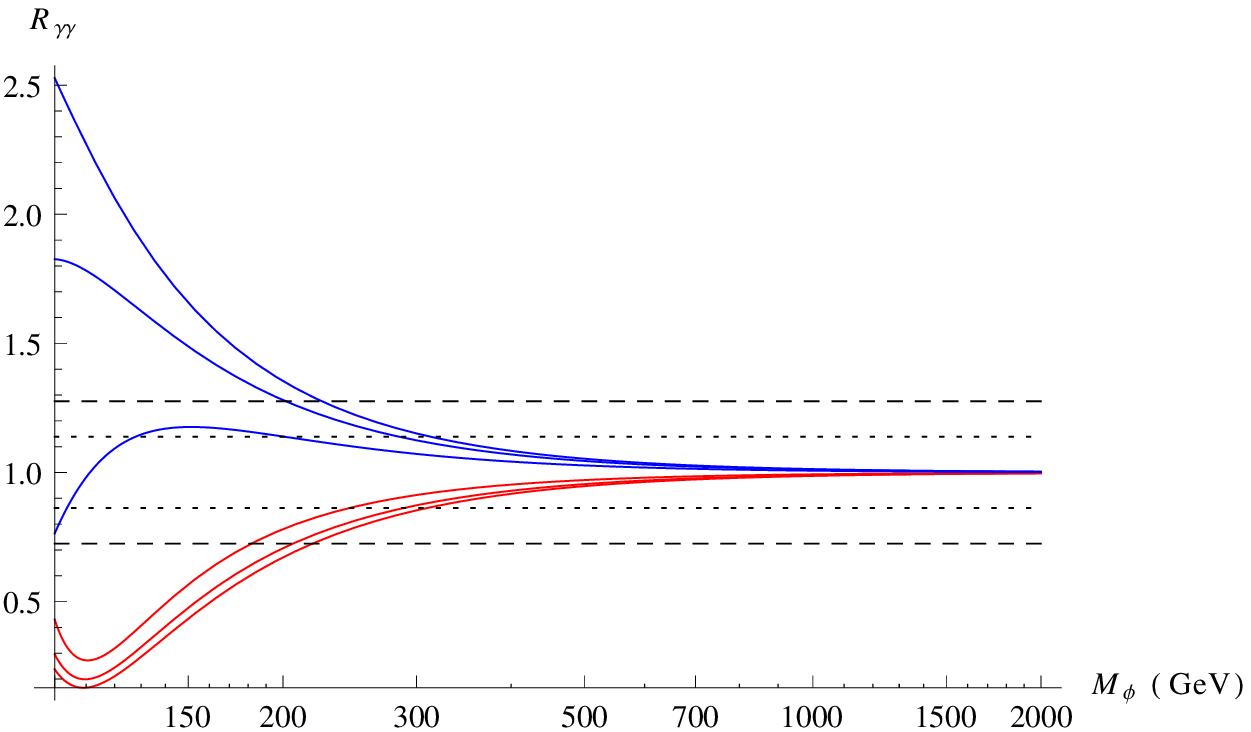}
\label{fig:subfig:QlC2}}

\subfloat[$|\lambda_\phi C(r_\phi)| = 1$]{
\includegraphics[width=3in]{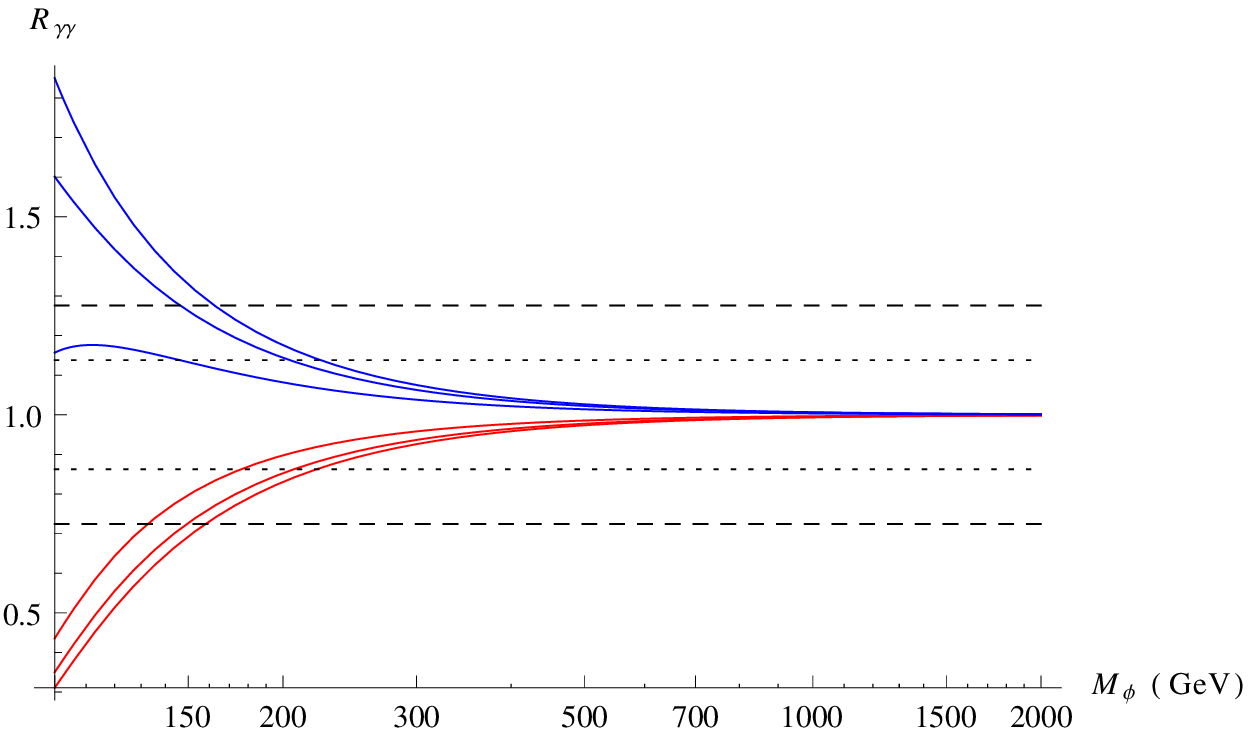}
\label{fig:subfig:QlC1}}
\hspace{0.2in}
\subfloat[$|\lambda_\phi C(r_\phi)| = 0.5$]{
\includegraphics[width=3in]{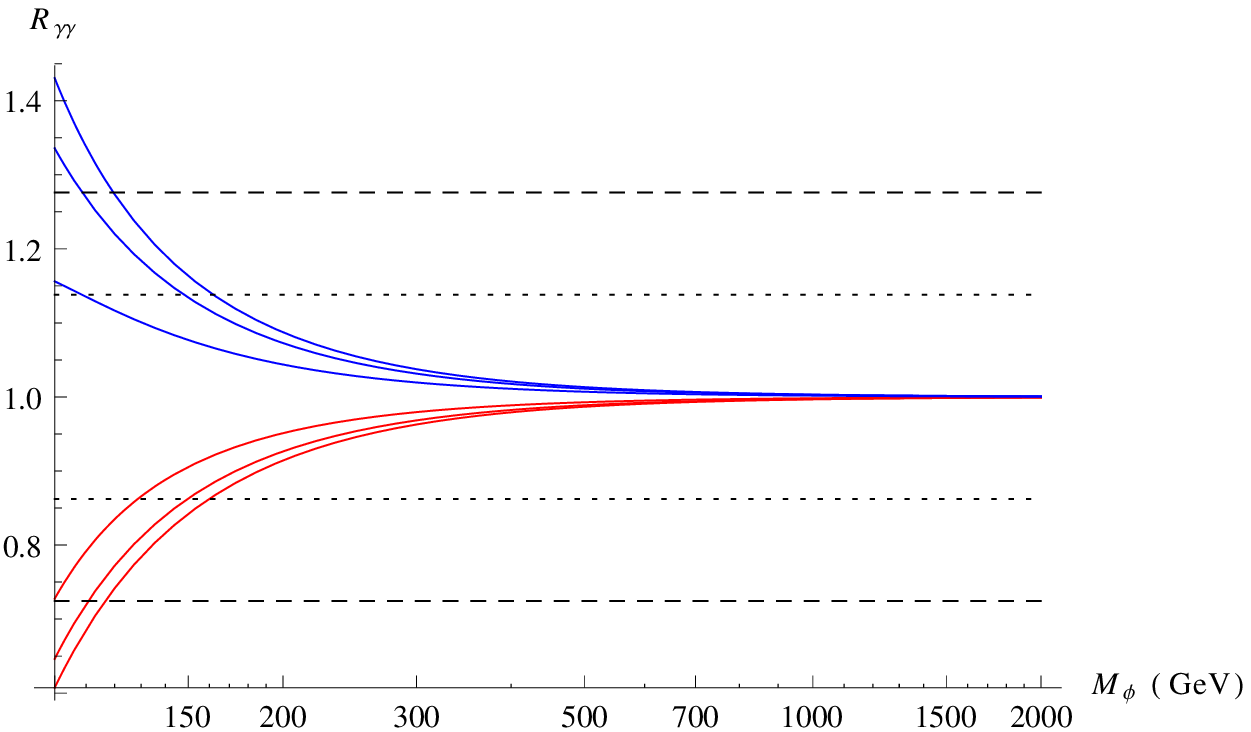}
\label{fig:subfig:QlC0p5}} \caption{The SM normalized production
rate $R_{\gm\gm}$ for the case of charged scalar NP for various
values of $\lambda_\phi C(r_\phi)$ and $\lambda_\phi
d(r_\phi)Q_\phi^2$. The blue (red) lines correspond to positive
(negative) values of $\lambda_\phi C(r_\phi)$. From top to bottom,
the blue lines correspond to $d(r_\phi)Q_\phi^2/C(r_\phi) =
1,2,4$. For red lines, this ordering is reversed. The dotted
(dashed) lines give bounds from the total theoretical $1\sigma$
($2\sigma$) uncertainties in the SM diphoton production rate.}
\label{fig:Raa}
\end{figure}
In Fig.~\ref{fig:Raa} we show how $R_{\gm\gm}$ depends on $\lambda_\phi C(r_\phi)$ and $\lambda_\phi d(r_\phi)Q_\phi^2$. We see for the mass range we consider, an enhancement in $R_{\gm\gm}$
would require in general a positive $\lambda_\phi$~\footnote{$C(r_\phi)$, $d(r_\phi)$, and $Q^2$ are all positive quantities}. Enhancement for negative $\lambda_\phi$ is possible, but happens only for
$|\lambda_\phi C(r_\phi)|$ sufficiently large, and in a much more restrictive mass range in the low mass region. In fact, the lower the mass, larger the enhancement. This is a general trend seen
in Figs.~\ref{fig:rxsGF} and~\ref{fig:RaaQ0}. The new feature here is that smaller the ratio $d(r_\phi)Q_\phi^2/C(r_\phi)$, larger the enhancement. This is because for $\lambda_\phi$ and
$C(r_\phi)$ fixed, increasing $d(r_\phi)Q_\phi^2$ depresses the diphoton width as is seen in Fig.~\ref{fig:rdwaa}. Note that the variation due to $d(r_\phi)Q_\phi^2$ is smaller when $\lambda_\phi$ is negative.

Since the $H \ra VV^*$ width is unchanged from the SM, and the
contribution of $\Gamma_{\gm\gm}$ to the total width is very
small, $R_{VV}$ is just $R_{\gm\gm}$ in the $Q_\phi = 0$ case
plotted in Fig.~\ref{fig:RaaQ0}. A difference between $R_{VV}$ and
$R_{\gm\gm}$ would favor the interpretation that colored scalars
are electrically charged. Comparing Fig.~\ref{fig:RaaQ0}
and~\ref{fig:Raa}, this appears to be the case in general.

\begin{figure}
\centering
\subfloat[$|\lambda_\phi C(r_\phi)| = 4$]{
\includegraphics[width=3in]{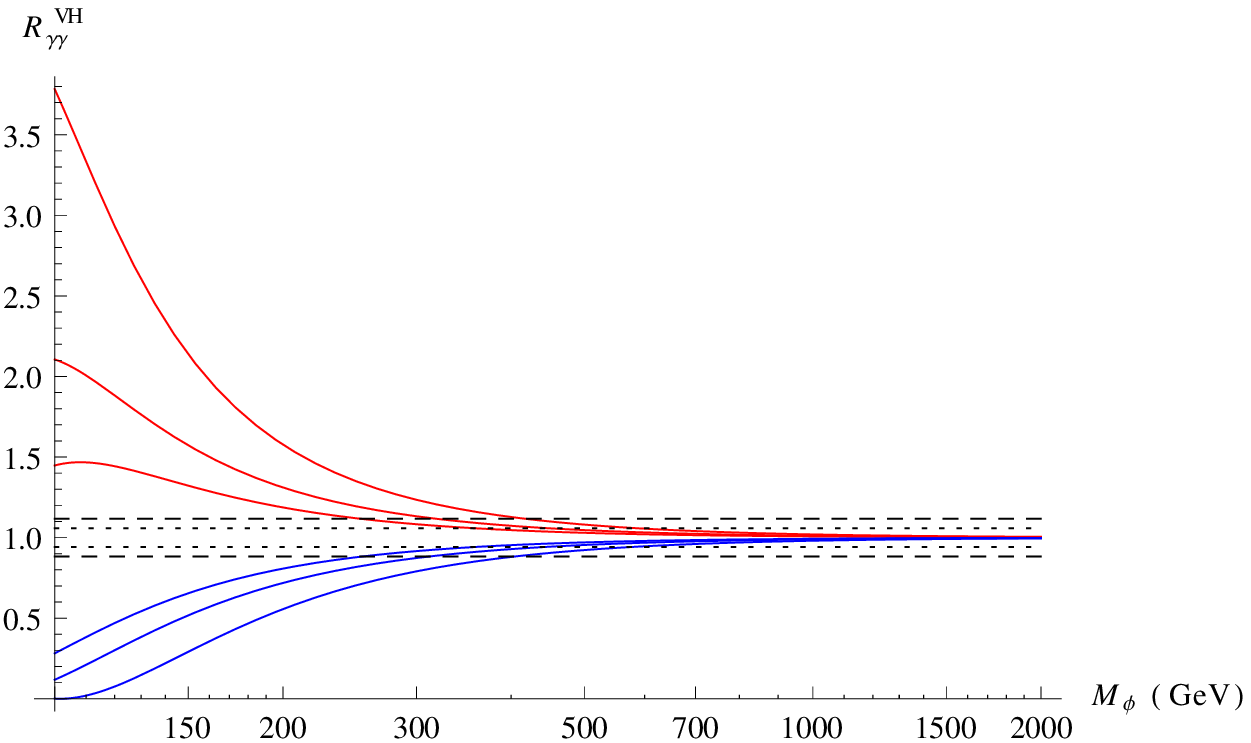}
\label{fig:subfig:VHlC4}}
\hspace{0.2in}
\subfloat[$|\lambda_\phi C(r_\phi)| = 1$]{
\includegraphics[width=3in]{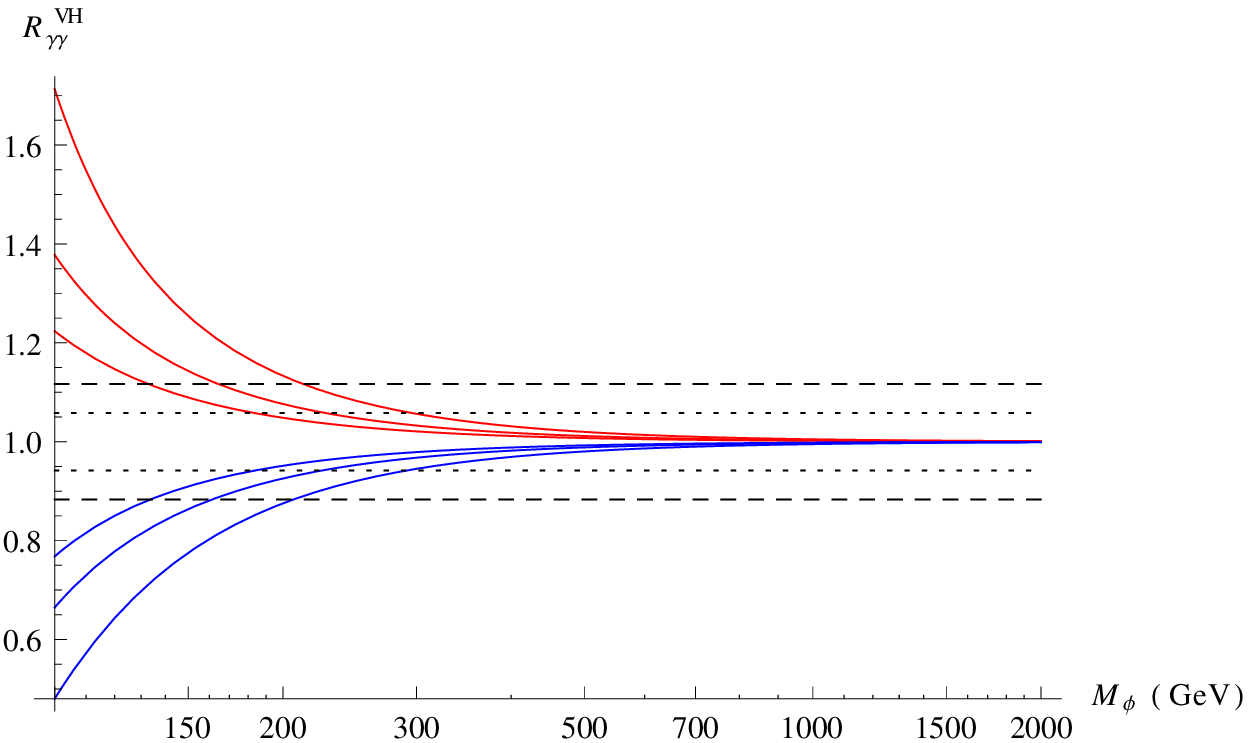}
\label{fig:subfig:VHlC1}} \caption{The SM normalized production
rate $R_{\gm\gm}^{VH}$ for various values of $\lambda_\phi
C(r_\phi)$ and $\lambda_\phi d(r_\phi)Q_\phi^2$. The blue (red)
lines correspond to positive (negative) values of $\lambda_\phi
C(r_\phi)$. From top to bottom, the blue lines correspond to
$d(r_\phi)Q_\phi^2/C(r_\phi) = 1,2,4$. For red lines, this
ordering is reversed. The dotted (dashed) lines give bounds from
the total theoretical $1\sigma$ ($2\sigma$) uncertainties in the
SM diphoton rate from VH production.} \label{fig:RVHaa}
\end{figure}
Another useful diagnostic here is $R_{\gm\gm}^{VH}$, which we show
in Fig.~\ref{fig:RVHaa}. The total theoretical $1\sigma$
uncertainty in the SM diphoton rate from VH production (the total
from both $WW$ and $ZZ$) is $\pm 5.8\%$. Note that
$R_{\gm\gm}^{VH}$ is just $\mathrm{Br}(H\ra\gm\gm)$ normalized to
the SM since the VH production is unchanged from the SM. Hence we
see enhancement (suppression) for negative (positive) values of
$\lambda_\phi C(r_\phi)$, like
$\Gamma_{\gm\gm}/\Gamma_{\gm\gm}^{SM}$ we see in
Fig.~\ref{fig:rdwaa}. Comparing $R_{\gm\gm}^{VH}$ with
$R_{\gm\gm}$ would allow the contribution from colored NP in
$\sigma_{gg}$ to be extracted.

\subsection{Scalars without color}
Models of color singlet scalars are very well studied for various
phenomenological motivations ranging from neutrino mass generation
to flavor physics; it is also an integral part of the minimal
supersymmetric SM (MSSM). We are interested here primarily in how
the 125~GeV Higgs signal at the LHC is relevant in these models.
The most important aspect for us is then the scalar potential. We
organize our study by classifying the scalars according to their
$SU(2)$ representation. Below we investigate in detail cases of
singlet, doublet, and triplet representations. Being color
singlets, the scalars alter only the Higgs diphoton decay width.
The crucial test here is to compare the branching ratios for the
$\gm\gm$ and the $VV^*$ channels, as deviation from the SM due to
NP is expected in the diphoton width.

\subsubsection{Singlets}
We begin with singlets $S$ without hypercharge. They are instrumental in constructing scalar dark matter models ~\cite{SDM} and hidden or shadow
Higgs models ~\cite{shhiggs}. A gauge invariant potential is given by
\begin{equation}
 -\frac{\mu_1^2}{2} H^\dagger H + \frac{\lambda_1}{4}(H^\dagger H)^4 -\frac{\mu_2^2}{2} S^\dagger S + \frac{\lambda_2}{4} (S^\dagger S)^2
 + \frac{\lambda_3}{2}(S^\dagger S)( H^\dagger H)
\end{equation}
If $S$ does not undergo SSB, it will not mix with the SM doublet $H$, and will thus have no effect on the LHC signal.

The more interesting case is when both $S$ and $H$ undergo SSB. The mass matrix of the two scalars is not diagonal and is given by
\begin{equation}
\frac{1}{2} \begin{pmatrix}
\lambda_1 v^2 & \lambda_3 v_s v \\
\lambda_3 v_s v & \lambda_2 v_s^2
\end{pmatrix} \,,
\end{equation}
where $v_s$ ($v$) denotes the VEV of the singlet (doublet). Since
$S$ is a SM singlet, the Higgs coupling will receive in every
vertex a universal suppression factor from the mixing angle that
arises from diagonalizing the mass matrix above. This implies that
$R_{\gm\gm}$ and $R_{VV}$ would be suppressed by the same factor.
An enhanced $R_{\gm\gm}$ would disfavor this case.

\subsubsection{Doublets}
As an archetypal example we consider here the two Higgs doublet
model (2HDM). For simplicity we assume CP invariance. To avoid
large flavor changing neutral currents, we assume also a discrete
$Z_2$ symmetry such that only one of Higgs doublets couples to
$u_R$, and the other $d_R$ and $e_R$. This is known as the Type-II
2HDM. The Higgs sector of the MSSM is a special case of this. A
general review is given in Ref.~\cite{2hdmr}.

The general gauge invariant scalar potential in this model is given by
\begin{align}\label{eq:2HDV}
V(\Phi_1,\Phi_2) =
&-\mu_1^2\Phi_1^\dagger\Phi_1 - \mu_2^2\Phi_2^\dagger\Phi_2 + \lambda_1(\Phi_1^\dagger\Phi_1)^2 + \lambda_2(\Phi_2^\dagger\Phi_2)^2
\notag \\
&+ \lambda_3(\Phi_1^\dagger\Phi_1)(\Phi_2^\dagger\Phi_2) - \lambda_4|\Phi_1^\dagger\Phi_2|^2 - \frac{\lambda_5}{2}\left[(\Phi_1^\dagger\Phi_2)^2 + (\Phi_2^\dagger\Phi_1)^2\right] \,.
\end{align}
Electroweak symmetry breaking is brought about by the vacuum expectation value (VEV) of $\Phi_i$. Explicitly we write
\begin{equation}
\Phi_1 =
\begin{pmatrix}
\phi_{1}^+ \\
\frac{v_1+h_1+i\chi_1}{\sqrt{2}}
\end{pmatrix} \,, \qquad
\Phi_2 =
\begin{pmatrix}
\phi_{2}^+ \\
\frac{v_2+h_2+i\chi_2}{\sqrt{2}}
\end{pmatrix} \,.
\end{equation}
The physics is best seen in the basis where only one of the doublets picks up a VEV:
\begin{equation}
\begin{pmatrix}
\Phi_1' \\
\Phi_2'
\end{pmatrix}
=
\begin{pmatrix}
 c_\beta & s_\beta \\
-s_\beta & c_\beta
\end{pmatrix}
\begin{pmatrix}
\Phi_1 \\
\Phi_2
\end{pmatrix} \,, \qquad
\langle\Phi_1'\rangle =
\begin{pmatrix}
0 \\
\frac{v}{\sqrt 2}
\end{pmatrix} \,, \quad
\langle\Phi_2'\rangle = 0 \,.
\end{equation}
We have used the shorthand notation $c_\theta = \cos\theta$, and we define $v^2 = v_1^2 + v_2^2$. The rotation angle is defined by $t_\beta\equiv\tan\beta = v_2/v_1$.

The physical degrees of freedom are projected out by a unitary
gauge transformation:
\begin{equation}
\Phi_{1}^\prime \ra
\begin{pmatrix}
0 \\
\frac{1}{\sqrt 2}(v + \eta)
\end{pmatrix} \,, \qquad
\Phi_{2}^\prime \ra
\begin{pmatrix}
H^+ \\
\frac{1}{\sqrt 2}(\phi + i\chi)
\end{pmatrix} \,.
\end{equation}
Here, $\eta$ and $\phi$ are scalar fields, while $\chi$ is a
pseudoscalar. The physical charged fields are the scalars $H^\pm$.
This is not yet the mass eigenbasis. For the neutral scalars, the
mass eigenbasis is given by
\begin{equation}
\begin{pmatrix}
H^0 \\
\Phi^0
\end{pmatrix}
=
\begin{pmatrix}
 c_\alpha & s_\alpha \\
-s_\alpha & c_\alpha
\end{pmatrix}
\begin{pmatrix}
\phi \\
 \eta
\end{pmatrix} \,.
\end{equation}
We identify the lighter state, $H^0$, to be the candidate Higgs
uncovered at the LHC. The mixing angle $\alpha$ is a complicated
function of the parameters in the scalar potential, and the
details are not needed here.

After some algebra the physical charged Higgs mass can be obtained from Eq.~\eqref{eq:2HDV} and is given
\begin{equation}\label{eq:chmass}
M^2_{H^\pm}\equiv\frac{1}{2}\bar{\lambda}v^2 = \half(\lambda_4 + \lambda_5)v^2 \,.
\end{equation}
The triple scalar coupling $H^0 H^+ H^-$ can be also be worked out:
\begin{align}\label{eq:3scalars}
\lambda_{H^0 H^+ H^-}&\equiv\lambda_s M_W \notag \\
&=
v\left\{\left[\half(\lambda_1+\lambda_2-\lambda_3+\bar{\lambda})s^2_{2\beta}+\lambda_3\right]s_\alpha
-\left[\lambda_1 s^2_{\beta}-\lambda_2 c^2_{\beta}+\half
(\lambda_3-\bar{\lambda})c_{2\beta}
\right]s_{2\beta}c_\alpha\right\} \,.
\end{align}
We see from above that the 2HDM is an example of non-decoupling scalars. If $M_{H^\pm}$ is taken large by taking $\bar{\lambda}$ large, $\lambda_{H^0 H^+ H^-}$ also becomes large.
Requiring that the theory be perturbative, all couplings should be at least less than $4\pi$, and $M_{H^\pm}$ cannot be much above the TeV scale.

Because of the scalar mixings, the couplings of the Higgs to fermions and gauge bosons are modified at the tree level. Thus $\sigma_0$ (in $\sigma_{gg}$) and $\Gamma_{\gm\gm}$ are modified in
addition to the charged Higgs contribution entering at the one-loop level.
\begin{table}[htbp]
\begin{tabular}{c|rr}
\multicolumn{1}{c|}{Vertex} & \multicolumn{1}{c}{$H^0_{2HDM}$} & \multicolumn{1}{c}{$H^0_{SM}$} \\
\hline
$\bar{t}t$ & $-i g_w\frac{M_t}{2M_w}\frac{\cos(\beta - \alpha)}{\sin\beta}$ & $-i g_w\frac{M_t}{2M_W}$ \\
$\bar{b}b$ & $i g_w\frac{M_b}{2M_W}\frac{\sin(\beta - \alpha)}{\cos\beta}$  & $-i g_w\frac{M_b}{2M_W}$ \\
$W^+ W^-$  & $i g^{\mu\nu}g_w M_W\sin\alpha$                                & $i g^{\mu\nu}g_w M_W$   \\
$ZZ$       & $i g^{\mu\nu}g_w\frac{M_Z}{\cos\theta_w}\sin\alpha$            & $i g^{\mu\nu}g\frac{M_Z}{\cos\theta_w}$
\end{tabular}
\caption{\label{table:hXX} Triple vertices of the Higgs to fermions and gauge bosons in 2HDM and the SM.}
\end{table}
In Table~\ref{table:hXX}, we list the modification to the relevant vertices. We also list their SM values for comparison. Note that we have included the $H^0 b\bar{b}$ coupling because if
$t_\beta$ is large, the bottom contribution cannot be neglected.

With the modified couplings in hand, $\sigma_0$ in the GF production cross-section is now
\begin{equation}\label{eq:2Hsig0}
\sigma_0 = \frac{G_\mu\alpha_s^2}{128\sqrt{2}\pi}\left|\half\left(\frac{c_\alpha}{t_\beta} + s_\alpha\right)F_{1/2}(\tau_t)
- \half(c_\alpha t_\beta - s_\alpha)F_{1/2}(\tau_b)\right|^2 \,,
\end{equation}
and the diphoton decay width
\begin{equation}\label{eq:2H2ga}
\Gamma_{\gm\gm} = \frac{G_\mu\alpha^2 M_H^3}{128\sqrt{2}\pi^3}\Bigg|s_\alpha F_1(\tau_W)
+ \frac{4}{3}\left(\frac{c_\alpha}{t_\beta} + s_\alpha\right)F_{1/2}(\tau_t) - \frac{1}{3}(c_\alpha t_\beta - s_\alpha)F_{1/2}(\tau_b)
+ \frac{\lambda_s}{\bar{\lambda}}F_0(\tau_{H^\pm})\Bigg|^2 \,.
\end{equation}
In deriving the above, we have expanded the trigonometric factors
in terms of $t_\beta$, and we have used Eqs.~\eqref{eq:chmass}
and~\eqref{eq:3scalars}.

Unlike the colored scalar case before, the Higgs decay widths can
be quite different from the SM one. If $\beta \simeq \alpha$, the
usual dominant $b\bar{b}$ decay channel can be suppressed by
$\sin(\beta - \alpha)$. In this case the $VV^*$ channel becomes
dominant. Currently, this does not appear to be what is observed,
although a firm conclusion is yet to be
reached~\cite{Atlas2,CMS2}. Assuming there are no accidental
cancellations in the parameters, the $b\bar{b}$ and the $VV$
widths are given by
\begin{equation}\label{eq:hwidth}
\Gamma_{b\bar{b}} = \left(c_\alpha t_\beta - s_\alpha
\right)^2\Gamma_{b\bar{b}}^{SM} \,, \qquad \Gamma_{VV} =
s_\alpha^2\Gamma_{VV}^{SM} \,.
\end{equation}

\begin{figure}
\centering
\subfloat[$\lambda_s/\bar{\lambda} = 1$.]{
\includegraphics[width=3in]{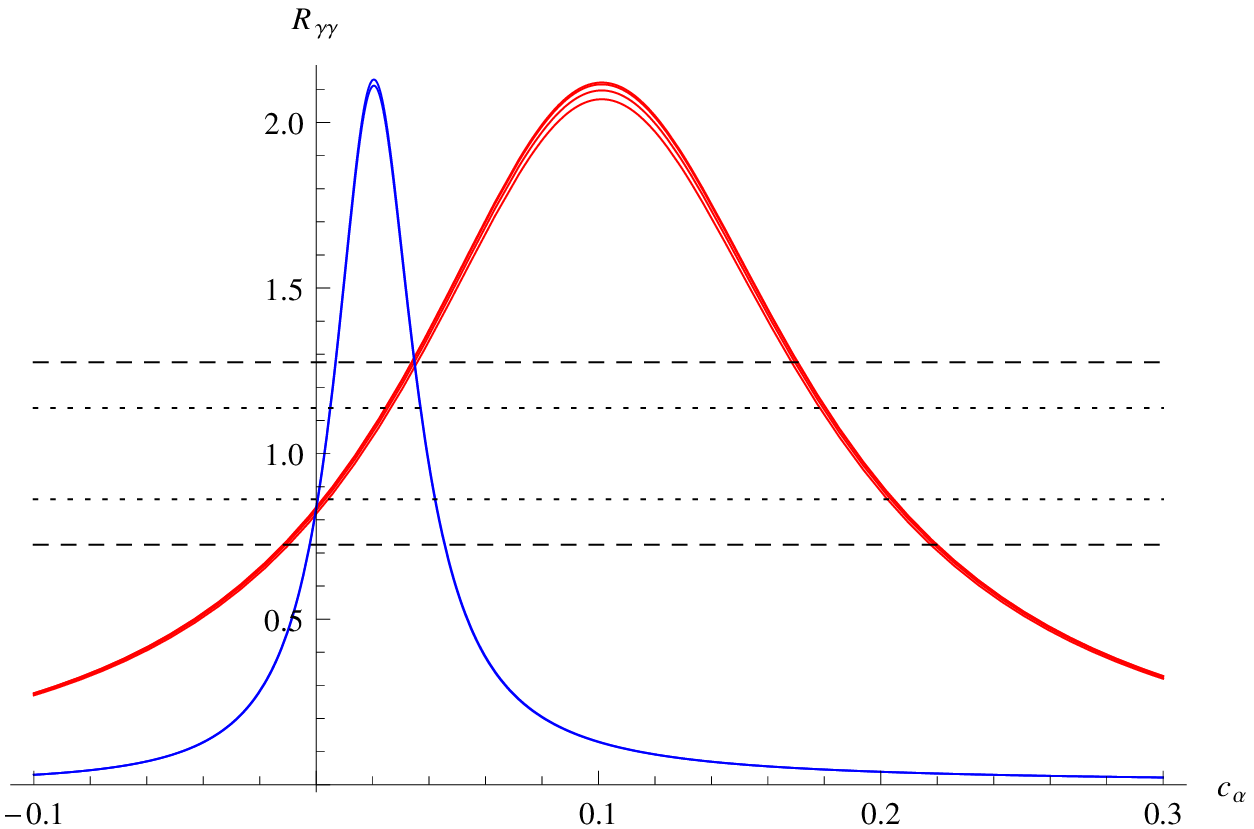}
\label{fig:subfig:tb20rl1}}
\hspace{0.2in}
\subfloat[$\lambda_s/\bar{\lambda} = -1$]{
\includegraphics[width=3in]{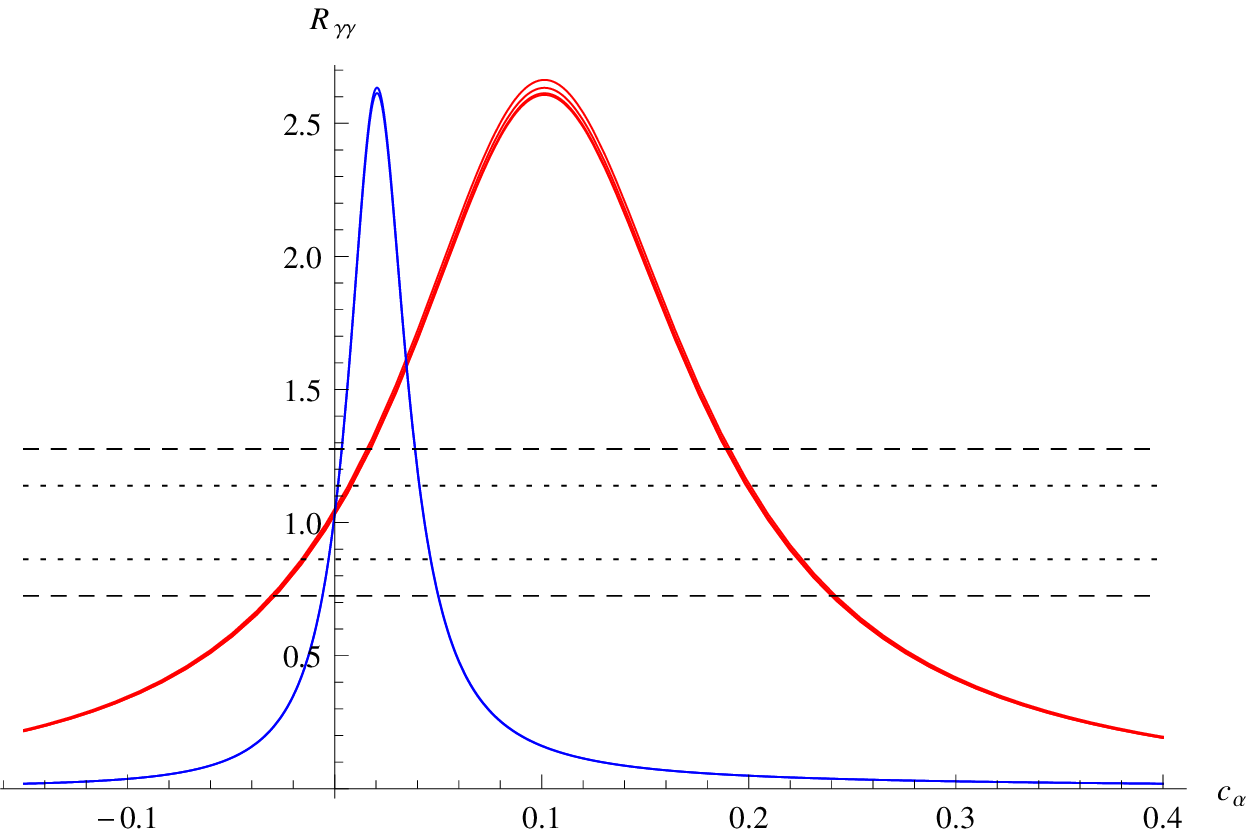}
\label{fig:subfig:tb5rl1}}

\subfloat[$\lambda_s/\bar{\lambda} = 10$]{
\includegraphics[width=3in]{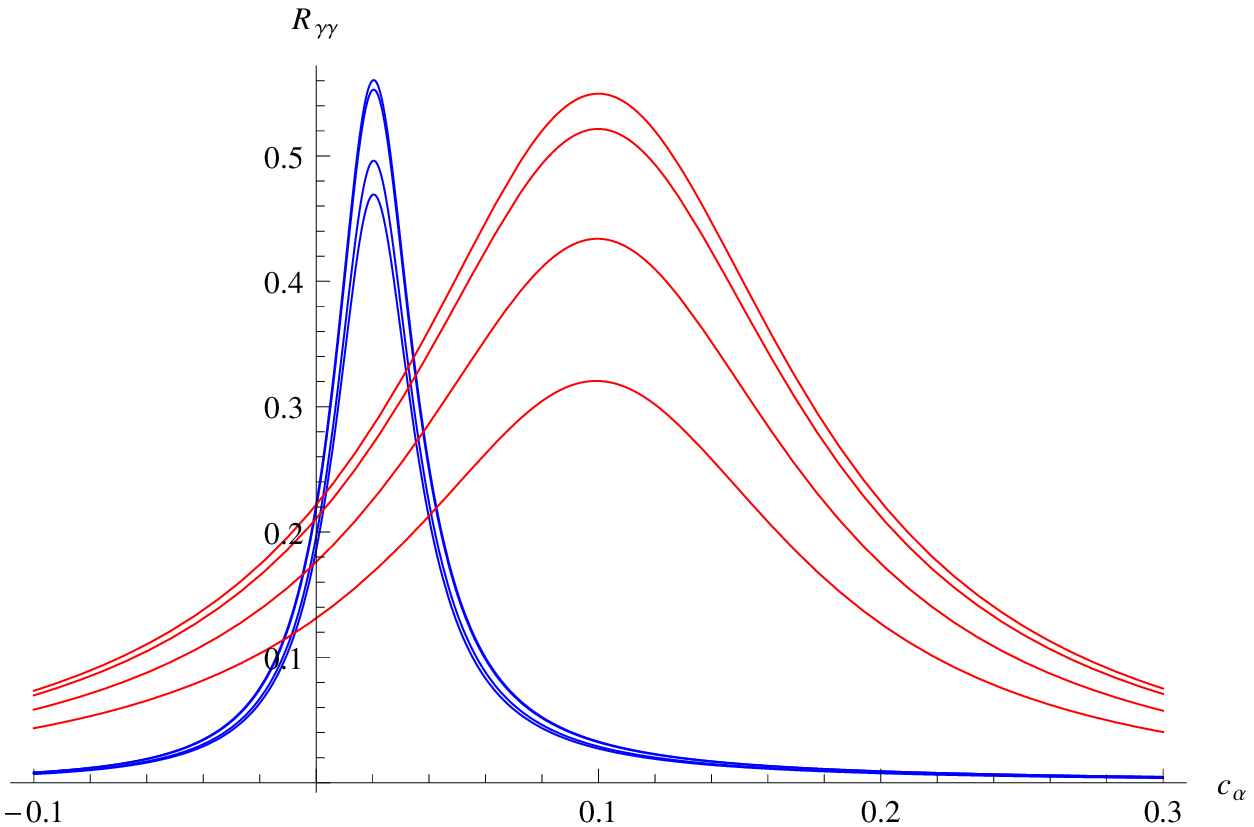}
\label{fig:subfig:tb20rl10}}
\hspace{0.2in}
\subfloat[$\lambda_s/\bar{\lambda} = -10$]{
\includegraphics[width=3in]{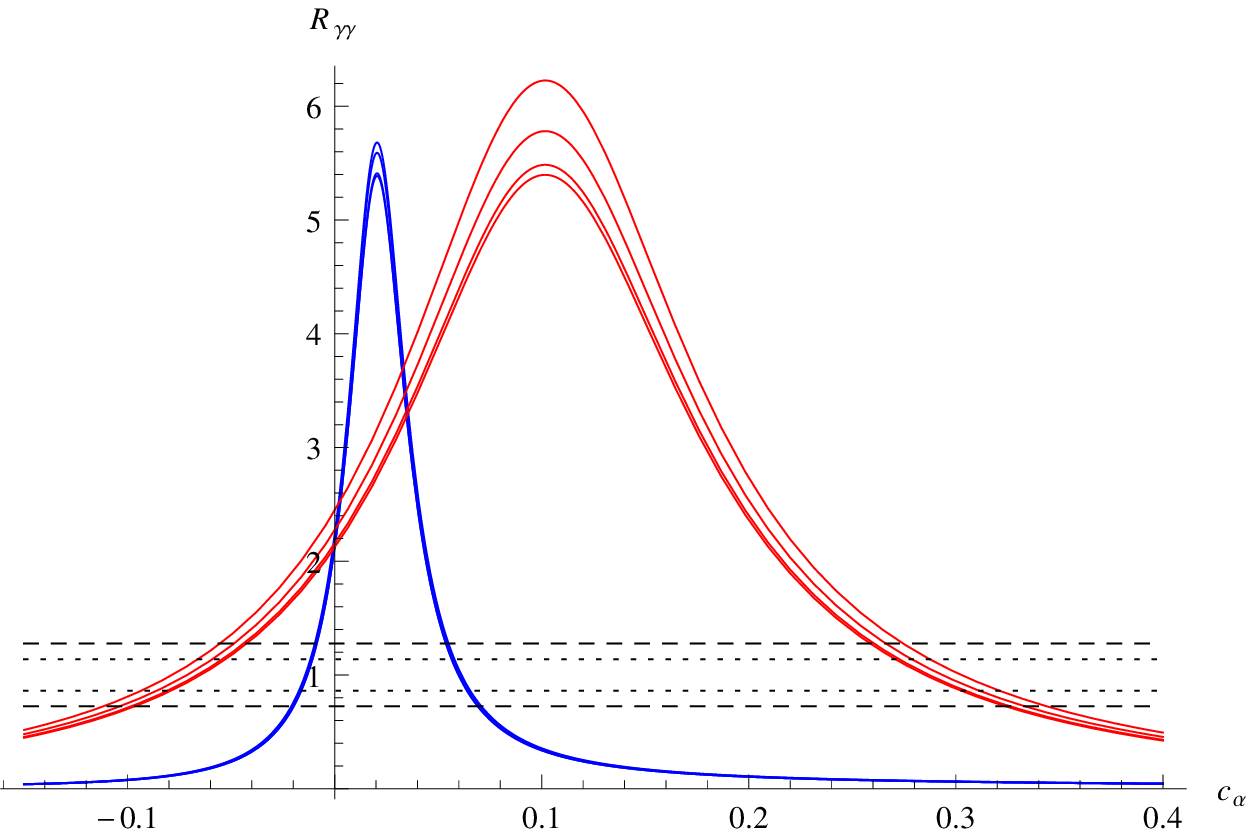}
\label{fig:subfig:tb5rl10}} \caption{The SM normalized production
rate $R_{\gm\gm}$ for various values of $\tan\beta$,
$\lambda_s/\bar{\lambda}$, and $M_{H^\pm}$. The red (blue) lines
correspond to $\tan\beta = 10\,(50)$. From bottom to top, the red
lines correspond to $M_{H^\pm} = 109.6,\,150,\,295,\,2000$~GeV,
and the blue $M_{H^\pm} = 168.1,\,198.5,\,548.1,\,2000$~GeV. The
orderings reverse for negative values of
$\lambda_s/\bar{\lambda}$. The dotted (dashed) lines give bounds
from the total theoretical $1\sigma$ ($2\sigma$) uncertainties in
the SM diphoton production rate.} \label{fig:RaaD}
\end{figure}
In Fig.~\ref{fig:RaaD} we show how $R_{\gm\gm}$ depends on the free parameters $t_\beta$, $\lambda_s/\bar{\lambda}$, and $M_{H^\pm}$. We see that there is little mass dependence in general for
$M_{H^\pm} > 500$~GeV, a manifestation of non-decoupling. The mass dependence also weakens as $\tan\beta$ increases, or as $\lambda_s/\bar{\lambda}$ decreases. In particular, for $\tan\beta = 50$
and $|\lambda_s/\bar{\lambda}| = 1$, $R_{\gm\gm}$ is virtually independent of $M_{H^\pm}$. 
For a given $t_\beta$ the peak in $R_{\gm\gm}$ is also independent
of $M_{H^\pm}$, as well as $\lambda_s/\bar{\lambda}$. The peak
shifts towards smaller $c_\alpha$ as $t_\beta$ increases. We see
that for large $\lambda_s/\bar{\lambda}$, large enhancement in
$R_{\gm\gm}$ can happen if it is negative, while suppression if
positive.

Experimentally, the process $B^+\ra\tau^+\nu$ provides a stringent bound on $t_\beta/M_{H^\pm}$.
Explicitly, we have~\cite{btaunu}
\begin{equation}
r_b = \frac{\mathrm{BR}(B^+\ra\tau^+\nu)}{\mathrm{BR}_{SM}(B^+\ra\tau^+\nu)} = \left(1-\frac {m_{B^\pm}^2 t_\beta^2}{M_{H^\pm}^2}\right)^2
\end{equation}
Taking the average of the BELLE~\cite{BELLE} and BABAR~\cite{BABAR} results, the Heavy flavor Averaging group~\cite{HFAG} finds the branching ratio $\mathrm{BR}(B^+\ra\tau^+\nu)$ to be
$(1.64\pm 0.34) \times 10^{-4}$. This implies that $r_b= 1.37\pm 0.39$. Using this at the 95\% confidence level, $M_{H^\pm} > 109.6$~GeV is allowed at
$t_\beta = 10$, while at $\tan\beta = 50$, $M_{H^\pm}$ is allowed between 168.1~GeV and 198.5~GeV, and also above 548.1~GeV.

For Type-II 2HDM, there is actually a $t_\beta$-independent bound of $M_{H^\pm} > 295$~GeV coming from the inclusive $b \ra s\gm$ decay~\cite{chmass}. However, this assumes that the only
NP contribution comes from the 2HDM and nothing else. Applying this bound would mean that in Fig.~\ref{fig:RaaD}, the lower (top) two red and blue curves in plots with positive (negative) $\lambda_s/\bar{\lambda}$ are ruled out.

\begin{figure}
\centering
\includegraphics[width=5in]{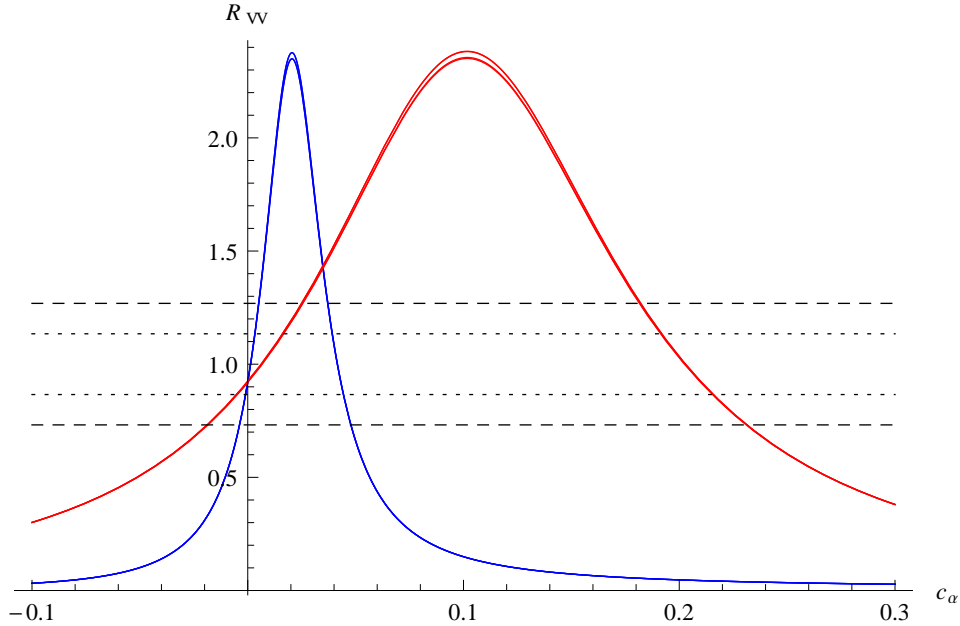}
\caption{The SM normalized production rate $R_{VV}$ for various
values of $\tan\beta$ and $\lambda_s/\bar{\lambda}$. The red
(blue) lines correspond to $\tan\beta = 10\,(50)$. The bottom
(top) lines correspond to $\lambda_s/\bar{\lambda} = -10\,(10)$.
The dotted (dashed) lines give bounds from the total theoretical
$1\sigma$ ($2\sigma$) uncertainties in the SM Higgs $VV^*$
production rate.} \label{fig:RVVD}
\end{figure}
In Fig.~\ref{fig:RVVD} we show how $R_{VV}$ depends on the free
parameters $t_\beta$, $\lambda_s/\bar{\lambda}$, and $M_{H^\pm}$.
The total theoretical $1\sigma$ uncertainties in the SM Higgs
$VV^*$ production rate is $\pm 13.4\%$. We see that $R_{VV}$ is
practically independent of $M_{H^\pm}$. This is because the mass
dependence enters only in $\Gamma_{\gm\gm}$ and is weak. Moreover,
the branching ratio to two photons is small compared with the
other channels. Thus the total width and hence $R_{VV}$ hardly
varies with $M_{H^\pm}$. For the same reason, there is also little
dependence on $\lambda_s/\bar{\lambda}$, which again enters only
in $\Gamma_{\gm\gm}$.

Comparing Fig.~\ref{fig:RaaD} and~\ref{fig:RVVD}, for a given $t_\beta$, $R_{VV}$ is quite different from $R_{\gm\gm}$ for large $\lambda_s/\bar{\lambda}$, but comparable for small
$\lambda_s/\bar{\lambda}$ ($|\lambda_s/\bar{\lambda}| \lesssim 1$). Thus a comparison between $R_{VV}$ and $R_{\gm\gm}$ may in addition be used to gauge the magnitude of
$\lambda_s/\bar{\lambda}$.

\subsubsection{Triplets}
Besides being of interest in their own rights, models incorporating both a doublet and a triplet Higgs fields are common in models of
neutrino masses from Type-II seesaw mechanism~\cite{trip} and quantum radiative corrections~\cite{radnu}.

We consider here the simple case where the Higgs triplet carries no hypercharge ($Y = 0$). The Higgs doublet and triplet fields are respectively given by
\begin{equation}
H=
\begin{pmatrix}
h^+ \\
\frac{1}{\sqrt{2}}(v_h + h^0 +i\chi)
\end{pmatrix} \,, \qquad
T=
\begin{pmatrix}
\half(v_T + T^0)       & \frac{1}{\sqrt{2}}T^+ \\
\frac{1}{\sqrt{2}}T^- & -\half(v_T + T^0)
\end{pmatrix} \,,
\end{equation}
where $v_h$ ($v_T$) denotes the VEV of the $H$ ($T$) field. For clarity, we have changed slightly the notation for the Higgs doublet from the previous subsection.

The most general, gauge invariant and renormalizable potential is
given by
\begin{equation}\label{eq:tripo}
V(H,T)=-\mu_{H}^2 H^\dagger H + \frac{\lambda_H}{4}(H^\dagger H)^2 -\mu_{T}^2\Tr T^2 + \frac{\lambda_T}{4}(\Tr T^2)^2 + \kappa\,H^\dagger H \Tr T^2 + \mu H^\dagger T H.
\end{equation}
The physical spectrum consists of pairs of neutral and charged scalars. An important feature here is that $v_T$ contributes to the $W$ boson mass,
whereas the $Z$ boson mass comes only from $v_h$. This gives rise to a tree level correction to the $\rho$ parameter:
\begin{equation}
\rho=1 + 4\frac{v_T^2}{v_h^2} \,.
\end{equation}
A global fit of the electroweak precision data gives $\rho =1.0008^{+.0017}_{-.0007}$~\cite{PDGrho}, from which we obtain the bound $v_T < 4$ GeV.

Parameters of the potential $V(H,T)$ are not all independent.
Minimizing the potential, we have relations:
\begin{align}
\mu_H^2 - \frac{1}{4}\lambda_H v_h^2 - \half\kappa v_T^2 + \half\mu v_T &= 0 \,, \\
\left(\mu_T^2 - \frac{1}{4}\lambda_T v_T^2 - \half\kappa v_h^2\right)v_T + \frac{1}{4}\mu v_h^2 &= 0 \,.
\end{align}
With $v_h$ and $v_T$ being the input parameters and non-zero, these imply
\begin{equation}\label{eq:kapmu}
\kappa = \frac{4\mu_T^2 - \lambda_T v_T^2}{v_h^2} - \frac{4\mu_H^2 - \lambda_H v_h^2}{2v_T^2} \,, \qquad
\mu = \frac{v_T(4\mu_T^2 - \lambda_T v_T^2)}{v_h^2} - \frac{4\mu_H^2 - \lambda_H v_h^2}{v_T} \,.
\end{equation}
Note that in the exact limit where $v_T = 0$, we can only minimize
the potential with respect to $H$, and we get the usual condition
$2\mu_H = v_h\sqrt{\lambda_H}$.

Consider now the charged states. From Eq.~\eqref{eq:tripo}, their mass matrix is given by
\begin{equation}
M_{\pm}^2 =
\begin{pmatrix}
\mu v_T & \half \mu v_H \\
\half \mu v_H & \frac{\mu v_{H}^2}{4 v_T}
\end{pmatrix} \,.
\end{equation}
Diagonalizing and the physical charged mass eigenstates are given
by
\begin{equation}
\begin{pmatrix}
H^\pm \\
G^\pm
\end{pmatrix}
=
\begin{pmatrix}
 c_\theta & s_\theta \\
-s_\theta & c_\theta
\end{pmatrix}
\begin{pmatrix}
T^\pm \\
h^\pm
\end{pmatrix} \,,
\end{equation}
where the mixing angle is defined by
\begin{equation}
c_\theta = \frac{v_h}{v} \,, \qquad s_\theta = 2\frac{v_T}{v} \,, \qquad v^2 = v_h^2 + 4v_T^2 \,,
\end{equation}
$G^\pm$ are massless would-be-Goldstone bosons, and $H^\pm$ the physical charged Higgs with mass given by
\begin{equation}\label{eq:MHpm}
M_{H^\pm}^2 = \mu v_T\left(1 + \frac{v_h^2}{4v_T^2}\right)
= (4\mu_T^2 - \lambda_T v_T^2)\left(\frac{1}{4} + \frac{v_T^2}{v_h^2}\right) - (4\mu_H^2 - \lambda_H v_h^2)\left(1 + \frac{v_h^2}{4v_T^2}\right) \,,
\end{equation}
after using Eq.~\eqref{eq:kapmu}. We see that since $t_\theta\equiv\tan\theta = 2(v_T/v_h)$ is small, $H^\pm$ are mostly composed of $T^\pm$.

From Eq.~\eqref{eq:MHpm}, the physical charged Higgs mass would be
naturally in the TeV range. If we take the limit $t_\theta \ll 1$
while keeping all other independent parameters ($\mu_{H,T}$ and
$\lambda_{H,T}$) fixed, $H^\pm$ would become very heavy and thus
decouple. However, if in addition we take $4\mu_H^2 - \lambda_H
v_h^2 = -k^2 v_T^2$ for some fixed $k$, then $M_{H^\pm} =
\frac{v}{2}\sqrt{\kappa + \frac{k^2}{2}}$ can remain at the weak
scale, and $H^\pm$ would not decouple. We note here that although
the two cases have different theoretical implication, the Higgs
signal itself would not be able to distinguish between the two.

Consider next the neutral states, whose mass matrix is given by
\begin{equation}
M_0^2 =
\begin{pmatrix}
\half\lambda_H v_h^2                  & \left(\kappa v_T - \half\mu\right)v_h \\
\left(\kappa v_T - \half\mu\right)v_h & \half\lambda_T v_T^2 + \frac{\mu v_h^2}{4v_T}
\end{pmatrix} \,,
\end{equation}
and the mass eigenbasis
\begin{equation}
\begin{pmatrix}
H^0 \\
\Phi^0
\end{pmatrix}
=
\begin{pmatrix}
 c_\xi & s_\xi \\
-s_\xi & c_\xi
\end{pmatrix}
\begin{pmatrix}
h^0 \\
T^0
\end{pmatrix} \,.
\end{equation}
The mass eigenvalues are given by
\begin{equation}
M_{H^0,\Phi^0}^2 = \frac{1}{8}\left(A\mp\sqrt{A^2 + 4B^2 t_\theta^2}\right) \,,
\end{equation}
where
\begin{equation}
A = 2(\lambda_H + \kappa)v_h^2 + 3\lambda_T v_T^2 - 4\mu_T^2 \,, \qquad
B = \lambda_T v_T^2 - 4\mu_T^2 \,.
\end{equation}
We take $H^0$ to be the lighter state and thus with the negative
sign. Expanding in small $t_\theta$ up to
$\mathcal{O}(t_\theta^3)$, the masses are given by
\begin{equation}
M_{H^0}^2 = \half\lambda_H v_h^2 - \frac{B^2 t_\theta^2}{4A} \,, \qquad
M_{\Phi^0}^2 = \frac{1}{4}(2\kappa v_h^2 + 3\lambda_T v_T^2 - 4\mu_T^2) + \frac{B^2t_\theta^2}{4A} \,,
\end{equation}
and the mixing
\begin{equation}
c_\xi = 1 - \frac{B^2}{2A^2}t_\theta^2 \,, \qquad s_\xi = \frac{B}{A}t_\theta \,.
\end{equation}
We see from this that $H^0$ is SM-like (mostly doublet), and we identify it as the LHC Higgs candidate.

We can now work out the cubic couplings of $H^0$ that contribute to its production and decay. The scalar coupling $H^0 H^+ H^-$ comes from the vertex $h^0 T^+ T^-$ in the gauge basis,
and is given by $c_\theta^2\,c_\xi\,\kappa v_h$. Since the triplet does not contribute to fermion mass generation, the $H^0 f\bar{f}$ Yukawa coupling is just the SM one with an additional
$c_\xi$ factor. For the $H^0$ coupling to gauge bosons, both the doublet and the triplet contribute. But since the former $\sim v_h$ while the latter $\sim v_T$, the triplet contribution
can be neglected in comparison to the doublet. Thus, the $H^0$ coupling to gauge bosons is again just the SM one with an extra $c_\xi$ factor. However, this factor cancels out in the
production rate between the production cross-section and the branching ratios. Thus, when calculating production rates, the modification from the SM come only from the charged Higgs
contribution to $\Gamma_{\gm\gm}$.

Working in the small $t_\theta$ limit with $(4\mu_H^2 - \lambda_H v_h^2)/v_T^2 = -k^2$ fixed, the $H^0 H^+ H^-$ coupling is simply
\begin{equation}
\lambda_{H^0 H^+ H^-} = \kappa v = \frac{4M_{H^\pm}^2}{v}\left(1 - \frac{k^2 v^2}{8M_{H^\pm}^2}\right) \,.
\end{equation}
It is convenient to redefine the coupling with $M_{H^\pm}^2$ scaled out, i.e. we define
\begin{equation}
\lambda_{H^0 H^+ H^-}M_W = \lambda\,g_w\,M_{H^\pm}^2 \,.
\end{equation}
The Higgs diphoton decay width is then
\begin{equation}\label{eq:CHga}
\Gamma_{\gm\gm} = \frac{G_\mu\alpha^2 M_H^3}{128\sqrt{2}\pi^3}\Bigg|F_1(\tau_W) + \frac{4}{3}F_{1/2}(\tau_t) + \lambda F_0(\tau_{H^\pm})\Bigg|^2 \,,
\end{equation}
and we see clearly that the charge Higgs do not decouple.

\begin{figure}
\label{fig:RaaTri}
\centering
\includegraphics[width=5in]{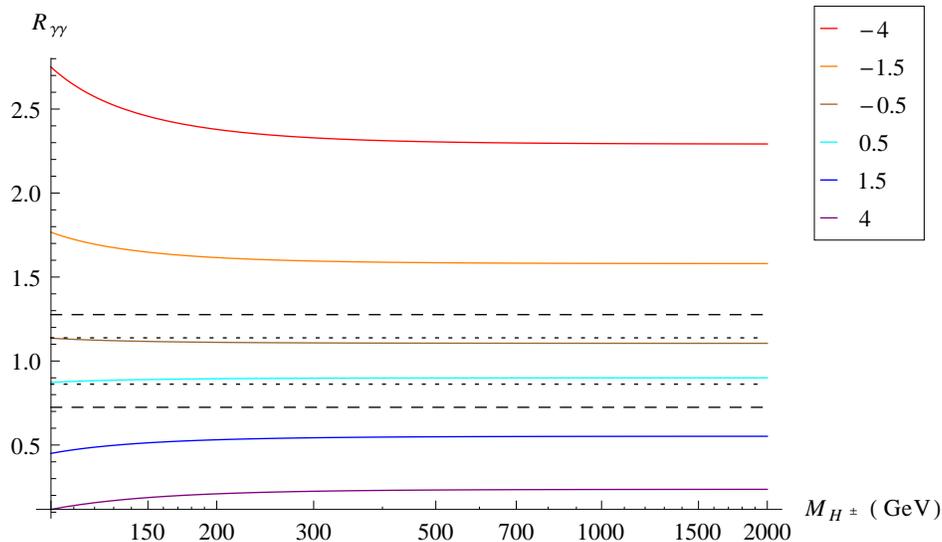}
\caption{The SM normalized production rate $R_{\gm\gm}$ for
various values of $\lambda$ as denoted by the colored lines. The
dotted (dashed) lines give bounds from the total theoretical
$1\sigma$ ($2\sigma$) uncertainties in the SM diphoton production
rate.} \label{fig:RaaTrip}
\end{figure}
We show in Fig.~\ref{fig:RaaTrip} the behavior of $R_{\gm\gm}$ as
a function of $M_{H^\pm}$ for various values of $\lambda$. We see
that the non-decoupling behavior sets in very quickly at about
300~GeV. Enhancement in $R_{\gm\gm}$ happens for negative values
of $\lambda$. Neglecting the effects of mixing which is small, the
predictions here are that $R_{\gm\gm} = R_{\gm\gm}^{VH} \neq 1$
and $R_{VV} = 1$.

\section{\label{sec:concl} Conclusions}
In this work, we have examined the constraints on scalar NP from
the recent LHC Higgs signals. We have studied in detail how
parameters relating to different new scalars can be constrained in
a general way by using $R_{\gm\gm}$, $R_{VV}$ and
$R_{\gm\gm}^{VH}$ in conjunction with each other. If the trend of
the current data persists, these constraints can be used to aid
the detection of new scalar states at the LHC, and even to
distinguish them. Indeed, the example of the Higgs triplet has
shown that its existence can be reveal by having measured
$R_{\gm\gm} = R_{\gm\gm}^{VH} \neq 1$ and $R_{VV} = 1$,
independent of what other roles it may play. Thus, the importance
of measuring $R_{VV}$ and $R_{\gm\gm}^{VH}$ in addition to
$R_{\gm\gm}$ cannot be overemphasized.

In general, if the mass generation mechanism of a new state is not
related to the electroweak symmetry breaking scale, decoupling
occurs. Examples we have given are colored scalars, with or
without electric charge. Their effect in the Higgs signal drop
away as they become heavier and heavier. For scalars from an
extended Higgs sector where mixings occur, this does not hold in
general. This has been illustrated in detail in our study of the
Type-II 2HDM and the $Y = 0$ triplet model. The SM normalized
production rates such as $R_{\gm\gm}$ would asymptote to constant
values different from unity in the large mass limit, and these
would give a measure of the effective triple scalar coupling. Of
course, non-decoupling behavior is already seen in fourth
generation studies. It is well known that fourth generation
fermion masses are proportional to their Yukawa couplings, and so
they are non-decoupling in our classification. Because of this,
even with the very limited statistics now, simple fourth
generation extension of the SM appears untenable~\cite{4thgen}.

We have paid special attention to the Type-II 2HDM because it is
archetypal in many theoretical and phenomenological constructions.
We have shown that by including constraints from B meson decays,
the Higgs signal can be used to probe the neutral scalar mixings,
which is a general feature of the model. For enhancement in
$R_{\gm\gm}$, small values of $c_\alpha$ are preferred in general.
For large $\tan\beta$, $R_{\gm\gm}$ is insensitive to the mass of
the charged Higgs. For smaller
values of $\tan\beta$, there is more sensitivity to the other
parameters.

As in all indirect searches of NP, it is possible that more than
one species of the new degrees of freedom can enter into the
observables, and accidental cancellations can happen to negate the
constraints found for a single species. This is a caveat we have
to bear in mind.

\begin{acknowledgments}
W.F.C. is supported by the Taiwan NSC, Grant
No.~99-2112-M-007-006-MY3. J.N.N. is partially supported by
Natural Science and Engineering Council of Canada. J.M.S.W. is
supported by NCTS.
\end{acknowledgments}

\end{document}